\documentclass[pre,showpacs,preprintnumbers,twocolumn,amsmath,amssymb,superscriptaddress]{revtex4-1}

\usepackage{algorithm}
\usepackage{algpseudocode}
\usepackage{booktabs}
\usepackage{moreverb}
\usepackage{graphicx}
\usepackage{latexsym}
\usepackage{graphics}
\usepackage{epsfig}
\usepackage{amsmath,amssymb,amsfonts,theorem,color,bm}
\usepackage{multirow}
\usepackage[normalem]{ulem}

\bmdefine\bu{\bm u} \bmdefine\bU{\bm U} \bmdefine\ba{\bm a}
\bmdefine\bb{\bm b} \bmdefine\bn{\bm n} \bmdefine\bk{\bm k}
\bmdefine\bx{\bm x} \bmdefine\by{\bm y} \bmdefine\bz{\bm z}
\bmdefine\bq{\bm q} \bmdefine\bQ{\bm Q} \bmdefine\bv{\bm v}
\bmdefine\bV{\bm V} \bmdefine\bw{\bm w} \bmdefine\bW{\bm W}
\bmdefine\bX{\bm X}
\bmdefine\be{\bm e} \bmdefine\bE{\bm E} \bmdefine\br{\bm r}
\bmdefine\bo{\bm 0} \bmdefine\boo{\bm o} \bmdefine\bOO{\bm O}
\bmdefine\bSigma{\bm \Sigma}
\bmdefine\btau{\bm \tau} \bmdefine\bh{\bm h} \bmdefine\bg{\bm g}
\bmdefine\bp{\bm p} \bmdefine\bP{\bm P}
\bmdefine\bs{\bm s}\bmdefine\bS{\bm S}
\bmdefine\bt{\bm t}
\bmdefine\bT{\bm T}

\bmdefine\bcalG{\bm{\mathcal G}} \bmdefine\bcalI{\bm{\mathcal I}}
\bmdefine\bcalT{\bm{\mathcal T}} \bmdefine\bcalL{\bm{\mathcal L}}
\bmdefine\bcalM{\bm{\mathcal M}} \bmdefine\bcalN{\bm{\mathcal N}}
\bmdefine\bcalR{\bm{\mathcal R}} \bmdefine\bcalS{\bm{\mathcal S}}
\bmdefine\bcalA{\bm{\mathcal A}} \bmdefine\bcalB{\bm{\mathcal B}}
\bmdefine\bcalC{\bm{\mathcal C}} \bmdefine\bcalD{\bm{\mathcal D}}
\bmdefine\bcalQ{\bm{\mathcal Q}} \bmdefine\bcalH{\bm{\mathcal H}}
\bmdefine\bcalU{\bm{\mathcal U}} \bmdefine\bcalV{\bm{\mathcal V}}
\bmdefine\bcalD{\bm{\mathcal D}}

\bmdefine\bPhi{\bm{\Phi}} \bmdefine\bvarphi{\bm{\varphi}}
\bmdefine\bSigma{\bm{\Sigma}} \bmdefine\bOmega{\bm{\Omega}}
\bmdefine\bF{\bm{F}} \bmdefine\bR{\bm{R}} \bmdefine\bbf{\bm{f}}
\bmdefine\bnabla{\bm{\nabla}}

\bmdefine\ip{\bm \cdot} \bmdefine\dip{\bm :}

\def\and{{\rm and}}

\newcommand{\beqn}{\begin{equation}}
\newcommand{\eeqn}{\end{equation}}

\makeatother

\begin{document}
	
	\title{Soliton approximation in continuum models of leader-follower behavior}
	\author{F.~Terragni}
\affiliation{Gregorio Mill\'an Institute for Fluid Dynamics, Nanoscience and Industrial Mathematics, Universidad Carlos III de Madrid, 28911 Legan\'{e}s, Spain}
\affiliation{Department of Mathematics, Universidad Carlos III de Madrid, 28911 Legan\'{e}s, Spain}
\author{W.~D.~Martinson}
	\affiliation{Wolfson Centre for Mathematical Biology, Mathematical Institute, University of Oxford, \\ Oxford OX2 6GG, United Kingdom}
	\author{M.~Carretero}
	\affiliation{Gregorio Mill\'an Institute for Fluid Dynamics, Nanoscience and Industrial Mathematics, Universidad Carlos III de Madrid, 28911 Legan\'{e}s, Spain}
\affiliation{Department of Mathematics, Universidad Carlos III de Madrid, 28911 Legan\'{e}s, Spain}
		
	\author{P.~K.~Maini\,}
\affiliation{Wolfson Centre for Mathematical Biology, Mathematical Institute, University of Oxford, \\ Oxford OX2 6GG, United Kingdom}
	\author{L. L. Bonilla$^*$}
\affiliation{Gregorio Mill\'an Institute for Fluid Dynamics, Nanoscience and Industrial Mathematics, Universidad Carlos III de Madrid, 28911 Legan\'{e}s, Spain}
\affiliation{Department of Mathematics, Universidad Carlos III de Madrid, 28911 Legan\'{e}s, Spain}
\affiliation{$^*$Corresponding author. E-mail: bonilla@ing.uc3m.es}

	\date{\today}

\begin{abstract}
Complex biological processes involve collective behavior of entities (bacteria, cells, animals) over many length and time scales and can be described by discrete models that track individuals or by continuum models involving densities and fields. We consider hybrid stochastic agent-based models of branching morphogenesis and angiogenesis (new blood vessel creation from pre-existing vasculature), which treat cells as individuals that are guided by underlying continuous chemical and/or mechanical fields. In these descriptions, leader (tip) cells emerge from existing branches and follower (stalk) cells build the new sprout in their wake. Vessel branching and fusion (anastomosis) occur as a result of tip and stalk cell dynamics. Coarse-graining these hybrid models in appropriate limits produces continuum partial differential equations (PDEs) for endothelial cell densities that are more analytically tractable. While these models differ in nonlinearity, they produce similar equations at leading order when chemotaxis is dominant. We analyze this leading order system in a simple quasi-one-dimensional geometry and show that the numerical solution of the leading order PDE is well described by a soliton wave that evolves from vessel to source. This wave is an attractor for intermediate times until it arrives at the hypoxic region releasing the growth factor. The mathematical techniques used here thus identify common features of discrete and continuum approaches and provide insight into general biological mechanisms governing their collective dynamics.
\end{abstract}
	\maketitle




\section{Introduction} \label{intro}
The interplay between discrete and continuum approaches informs our understanding of many biological processes, such as morphogenesis \cite{sek03,fri09,och12,alt17,han17,li19,vol18,wei09}, aggregation and swarming \cite{oku86,vic12,ber13,ber16,cav18,gon23}, pattern formation \cite{cat10,sme16}, bacterial motion \cite{igo01,igo04}, tissue repair \cite{pou07,bru14,rav15}, tumor invasion and metastasis \cite{fri95,fri03,fri09,gan20}. These phenomena all display elements of collective behavior, in which groups adopt unique behaviors not observed in smaller numbers of individuals. This comprises a central area of interest for soft and active matter physics, but collective cell behavior has the  additional complexity that cell groups have the ability to adopt a wide variety of fluid-like, solid-like, or even glass-like states by undergoing so-called flocking and jamming transitions \cite{ang11,par15,bi16,mal17,pal19}. Consequently, the mechanisms underlying collective phenomena remain poorly understood in general. Some insight may be provided by mathematical modelling, as it provides an abstract setting in which to evaluate different hypotheses. Two approaches are largely used to represent cells in a collective. One approach, known as discrete modelling, involves tracking the evolution of each individual cell \cite{hak17,tre18}. Due to the ability of these discrete approaches to represent each member of a collective, many biological mechanisms can be straightforwardly incorporated into discrete approaches and be directly tested in the laboratory. However, the long-time behavior of these models is usually difficult to ascertain without extensive and costly computation. This motivates the second type of modelling approach, which involves representing the whole population as a continuous function that evolves in space and time according to a set of partial differential equations (PDEs). While continuum approaches typically describe the ensemble average behavior of a collective, and hence cannot be used in general to resolve individual cells, these models are much faster to simulate, more amenable to analysis, and can provide insight into the important mechanisms governing the phenomenon of interest. For processes spanning many length and time scales, a combination of discrete and continuum approaches can be particularly useful. 

An important example of a biological phenomenon in which mathematical modelling has helped uncover important underlying mechanisms is angiogenesis, the process by which new blood vessels grow from existing vasculature. This complex multiscale process is the basis of organ growth and regeneration, tissue repair and wound healing in healthy conditions \cite{car05,CJ11,GG05,fruttiger,CT05,pot11,szy18}. Disruptions to the natural balance of pro- and anti-angiogenic factors, by contrast, are linked with various pathological diseases such as cancer, diabetes, and retinopathies \cite{pot11,fol71,byr10,saw17,pau17,veg21}. Angiogenesis is triggered by hypoxic (oxygen-lacking) cells that secrete diffusible growth factors which travel to nearby primary blood vessels. The binding of these growth factors to endothelial cells lining the primary vessel causes the latter to detach, move towards the hypoxic region, and build capillaries which transport blood, oxygen, and nutrients. A growing capillary is led by so-called ``tip'' cells that sense and move up the gradient of growth factors, in a process known as chemotaxis, to reach the hypoxic region. Stalk cells proliferate along the path of tip cells and construct the nascent capillary. A tip cell may encounter another tip cell or growing capillary during the course of migration; when it does so, it fuses with the object in a process called ``anastomosis'' which results in the new vessel forming a closed loop that supports blood flow. Tip cells may also emerge along the length of capillaries, which enables the creation of multiple branches in the new network.

Mathematical models of angiogenesis capture these dynamics of cell movement, branching, and anastomosis by describing tip and stalk cells as leaders and followers, respectively. In addition to angiogenesis, such leader-follower frameworks are important in explaining aspects of morphogenesis \cite{och12,alt17,li19,vol18} and wound healing \cite{ome03,sep13,bon20plos}. Multiple types of frameworks have been constructed to describe angiogenesis, ranging from continuum approaches described by PDEs \cite{lio77,bal85,cha93,byr95,cha95,and98} to discrete approaches using agent-based models (ABMs) \cite{veg21,bau07,bau09,ben08,an09,lie15,osb17,met19,ren20,veg20,met19,ste21,jaf21}, mesoscale approaches relying on tools from kinetic theory \cite{bel15}, or hybrid approaches combining aspects of discrete and continuum frameworks to simulate cells and their microenvironment \cite{cap09,rej11,tra11}. One important discrete model for angiogenesis, for instance, examined minimal mechanisms that could lead to the branched vessel networks resembling those observed in vivo \cite{and98}.  Continuum models of angiogenesis have also quantified how chemotaxis and branching determined the speed and distribution of tip cells \cite{byr95}. For further information about mathematical modelling of angiogenesis, we refer to the following reviews \cite{and98,pla03,man04,qut09,sci13,vil14,hec15,spi15,vil17,per17,bon19,fle20}. 

One drawback of many continuum models used to simulate angiogenesis is that the equations are constructed with a phenomenological ``top-down'' approach, which involves deducing the PDEs through principles such as the conservation of mass, energy, etc. For example, so-called ``snail-trail'' models consider non-linear stalk cell proliferation along tip cell trajectories \cite{bal85,byr95,hec15, pet96, con15} that are inspired by mathematical frameworks used to study branching patterns in fungal growth \cite{ede82}. Consequently, these continuum models can be difficult to link to the underlying biology and can be deceptively difficult to analyze, even though they form a mathematically interesting paradigm for leader-follower behavior. This motivated the derivation of ``coarse-grained'' PDEs, which can be obtained by investigating the ensemble average behavior of cell-based angiogenesis models using techniques from statistical mechanics \cite{ish17,her21,tri23,bon14,spi15,alb07,bak10,mar13,ter16,bon16,bon16pre,pil17,mar20,bon20,mar21}. While different ABMs can lead to quite different continuum equations \cite{ter16,pil17}, methods from asymptotic analysis \cite{mar21} show that there are parameter regimes for which these different continuum models produce identical dynamics at leading order. These ``leading order'' PDEs (LO-PDEs) suggest a shared set of mechanisms that are inherent to the leader-follower dynamics exhibited in angiogenesis and admit traveling wave solutions of time-varying amplitude, which in certain cases may be approximated by self-similar solutions.

Our main result in this paper is that the LO-PDEs in simple geometries have soliton-like solutions with slowly varying amplitudes, similar to those observed in \cite{bon16}. We consider a 2D scenario in which the primary blood vessel emitting tip cells and the hypoxic regions are situated along separated parallel vertical lines, such that we may average the PDE along the vertical direction to obtain a 1D approximation. Numerical simulations of the resulting LO-PDEs show that solitons are attractors for intermediate times until they arrive at the neighborhood of the hypoxic region. Soliton-like solutions were previously found in continuum PDE descriptions of hybrid stochastic angiogenesis models \cite{bon16,bon16pre,bon20}.

The rest of the paper is as follows. In Section \ref{sec:2}, we describe the leading order dynamics of different coarse-grained discrete angiogenesis models and its relation to a hybrid stochastic model. In Section \ref{sec:3}, we describe the approximation of the numerical solutions of the LO-PDEs by a soliton-like wave. We derive the collective coordinate equations (CCEs) that govern the shape and velocity of the soliton. For a given simple linear profile of the tumor angiogenic factor (TAF), we find in Section \ref{sec:4} that the soliton position is well approximated by the CCEs but its shape is not. This shortcoming stems from ignoring the transversal modulation of the TAF profile in a 2D  setting. Section \ref{sec:5} shows that the CCEs accurately predict the shape and motion of the soliton for a quasi-steady Gaussian TAF profile. This is so after a short transient formation stage and until the soliton arrives at the tumor. Increasing the distance between the primary vessel and tumor enlarges the time interval over which the CCEs provide accurate approximations to the soliton dynamics, as shown in Section \ref{sec:6}. In Section \ref{sec:7}, we discuss the effect of tip-to-tip anastomosis on soliton  evolution. Lastly, Section \ref{sec:8} is devoted to concluding remarks.


\section{Continuum model} \label{sec:2}

\subsection{Leading order equations}
\begin{figure}[h!]
	\begin{center}
		\includegraphics[width=8cm]{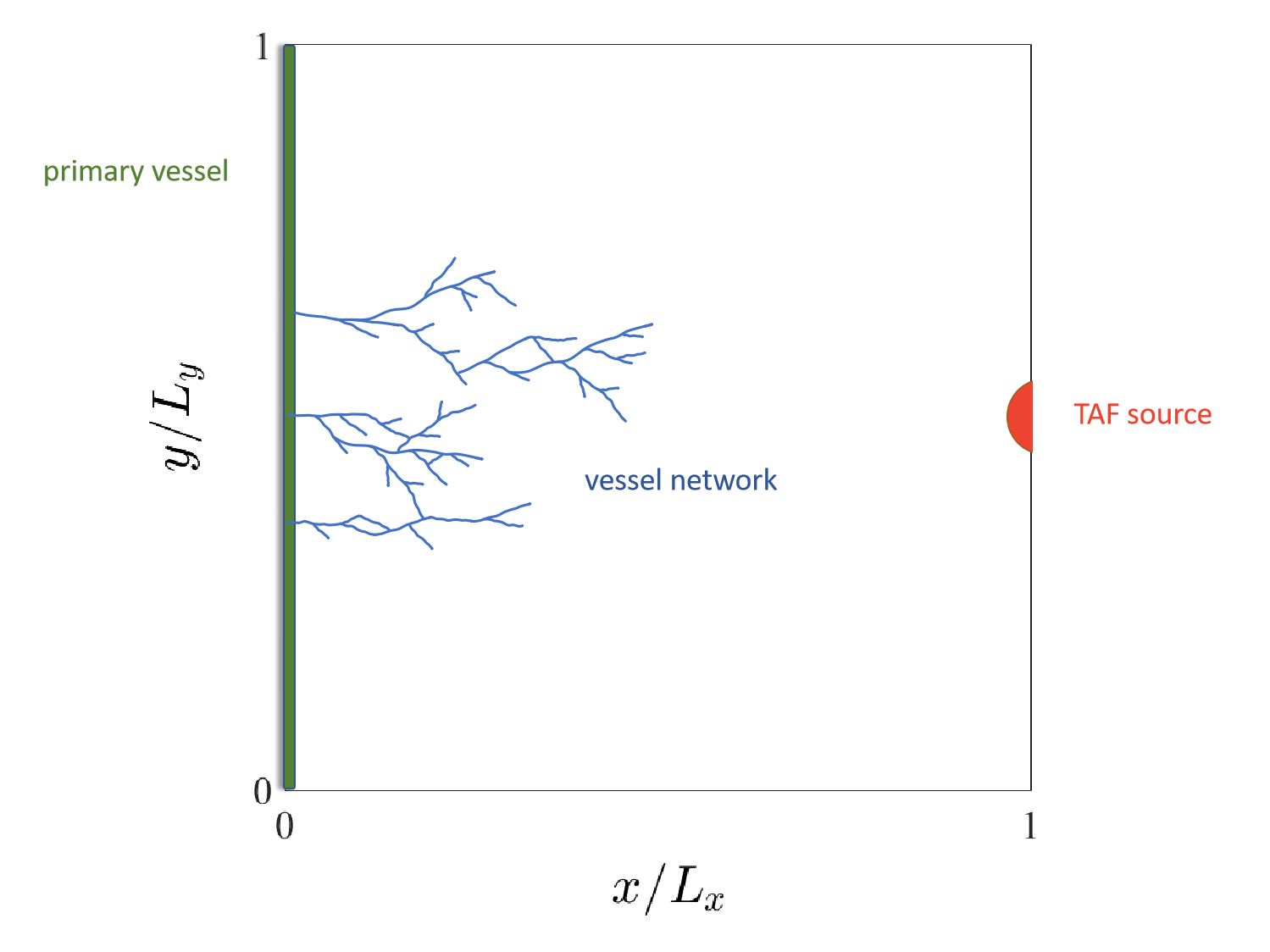}
	\end{center}
	\vskip-0.5cm
	\caption{Sketch of the geometry showing how blood vessels sprout from the primary vessel at $x=0$ and move to the source of TAF at $x=L_x$.\label{fig:0}}
\end{figure}

We consider the leading order angiogenesis model derived in \cite{mar21}, under the assumptions
of chemotaxis-dominated tip cell movement and relatively low branching rates. This two-dimensional (2D) continuum model is
described by the following dimensionless coupled LO-PDEs
\begin{subequations}\label{eq1}
\begin{eqnarray}
	\frac{\partial N}{\partial t} &=& D\nabla^2 N - \chi \nabla \cdot \!\left(N \nabla C\right)\! + \lambda N \frac{C}{1+C}\nonumber\\
	&-& \mu \,a_e N E - \mu \,a_n N^2, \label{eq1a} \\
	\frac{\partial E}{\partial t} &=& \mu \,N, \label{eq1b}
\end{eqnarray}\end{subequations}
for $\mathbf{x} = (x,y)$, with $0 < x < L_x$ and $0 < y < L_y$. The primary vessel and the tumor are located at $x=0$ and $x=L_x$, respectively 
(see Fig.~\ref{fig:0} for a schematic cartoon of this setup). Here  $C(\mathbf{x},t)$, $N(\mathbf{x},t)$, and $E(\mathbf{x},t)$ denote the TAF concentration, the density of tip cells, and the density of stalk cells, respectively. The positive parameter $D$ is the diffusion coefficient of tip cells, and corresponds to the influence of random movement,  $\chi$ is the chemotactic sensitivity of tip cells, $\lambda$ is the rate at which branching of new sprouts occurs, and $\mu$ is a baseline rate of anastomosis that is further modulated by the values of $a_e$ and $a_n$, which denote the specific rates of tip-to-sprout and tip-to-tip anastomosis, respectively. 

Tip and stalk cell densities are driven by a 2D TAF field (see below and Fig.~\ref{fig:0}). However, all terms depending on $C$ in Eq.~\eqref{eq1a} can be reduced to one spatial variable by column averaging (namely, averaging in the $y$-direction) in the same fashion as discussed in \cite{mar21}. The different terms on the right-hand-side of Eq.~\eqref{eq1a} describe random diffusion, chemotaxis, tip branching, tip-to-sprout anastomosis, and tip-to-tip anastomosis, while the time evolution of stalk cells is given by a production term depending on tip cell density according to Eq.~\eqref{eq1b}. The LO-PDE \eqref{eq1b} was derived in \cite{mar21} and it does not include flux of tip cells and production of stalk cells due to anastomosis \cite{pil17}, which are of higher order. The branching term in Eq.~\eqref{eq1a} saturates as $C\to\infty$, which is more realistic than the term linear in $C$ used in Ref.~\cite{mar21}. In this paper, we want to describe the evolution of traveling waves until they arrive near the hypoxic region, but not their interaction with the latter. Thus, we impose no-flux boundary conditions for the tip cell density  \cite{mar21}, and do not study the interaction of the waves with the boundary. Such a study would require using more realistic boundary conditions at $x=L_x$. We impose the same non-negative functions as in \cite{mar21} to represent the initial conditions for $N$ and $E$. \\

\begin{figure}[h!]
	\begin{center}
		\includegraphics[width=8cm]{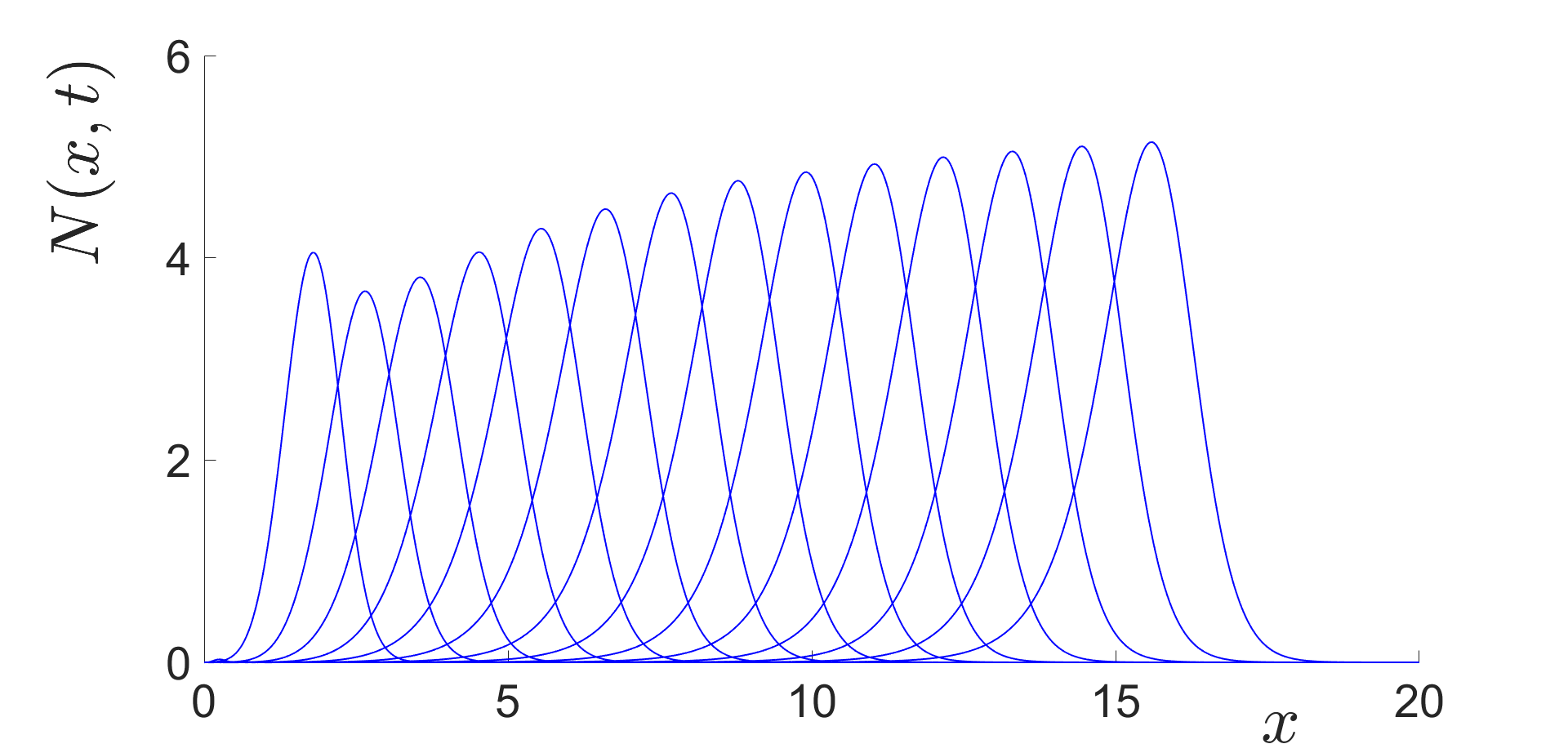}
	\end{center}
	\vskip-0.5cm
	\caption{Snapshots of the tip cell density taken at times $4+j(\Delta t)$ (with $\Delta t = 2$) between $t = 4$ and $t = 30$, $j=0,1,\ldots,13$. They have been numerically
		computed by means of Eqs.~\eqref{eq2}, assuming a quasi-steady, 1D
		linear TAF concentration $C(x) = x$ and $L_x = 20$. The scale on the vertical axis
		is $\times 10^{-5}$. \label{fig:1}}
\end{figure}

The column-averaged PDEs in one spatial variable derived from Eqs.~\eqref{eq1}, together with the boundary and initial conditions, are
\begin{subequations}\label{eq2}
\begin{eqnarray}
&&	\frac{\partial N}{\partial t} = D\frac{\partial^2 N}{\partial x^2} - \chi \frac{\partial}{\partial x}\!\left( N \frac{\partial C}{\partial x}\right)\! + \lambda N \frac{C}{1+C}\quad\,\nonumber\\
	&&\quad\quad - \mu \,a_e N E - \mu \,a_n N^2, \label{eq2a} \\
&&	\frac{\partial E}{\partial t} = \mu \,N, \label{eq2b}\\
&&	D\frac{\partial N}{\partial x}-\chi N\frac{\partial C}{\partial x}=0\quad\mbox{at } \,x=0,\,  L_x,\label{eq2c}\\
&&	N(x,0)=G(x),\quad E(x,0)=H(x). \label{eq2d}
	\end{eqnarray}\end{subequations}
Strictly speaking, column averaging gives a solution equivalent to that of the 2D equations only when the TAF field does not vary in the direction transversal to the travelling front. However, the authors found in \cite{mar21} that there were some situations in which the TAF field did vary in the $y$-direction but the numerical solution of the column averaged solution accurately represented that of the full 2D model. Numerical simulations of Eqs.~\eqref{eq2} are performed after discretizing the spatial derivatives with centered finite differences on a uniform mesh and by using the MATLAB solver \texttt{ode15s} for time integration. Figure \ref{fig:1} shows the tip cell density evolution over time for $L_x = 20$ and the following set of parameter values consistent with chemotaxis dominated transport, small diffusion and relatively low branching rate \cite{bon16pre}: $D = 0.04$, $\chi = 0.24$, $\lambda = 0.73$,
$\mu = 236$, $a_e = 0.14$, and $a_n = 0$. These values will be used throughout the paper,
unless otherwise stated.
In this case, a source of TAF is assumed to be located at $x = L_x$, considering a 
quasi-steady, one-dimensional (1D) linear concentration $C(x) = x$.
A wave is generated at $x = 0$ and travels forward in the direction of the TAF gradient.
Indeed, this is a 1D continuum, macroscopic description of the underlying
stochastic process detailed in \cite{mar21} and references therein.
Tip cells sprout from a primary vessel at $x = 0$ and migrate towards the right,
attracted by the TAF source. They randomly branch and anastomose, thus generating a vascular network. \\[3mm]

\subsection{Hybrid stochastic tip cell model equations}
From the hybrid stochastic tip cell model of \cite{cap09,bon14}, we can track the density of active tip cells, $p(\mathbf{x},\mathbf{v},t)$, and the TAF concentration, $C(\mathbf{x},t)$, by ensemble averages over realizations of the stochastic process \cite{ter16}. Active tip cells are those moving or branching out at a given time. When an active tip cell meets the trajectory of another tip cell, it anastomoses, stops there, and ceases to exist. Then, we may derive the following nondimensional PDEs \cite{ter16,bon18} 
\begin{widetext}
\begin{subequations}\label{eq3}
\begin{eqnarray} 
\frac{\partial}{\partial t} p(\mathbf{x},\mathbf{v},t)\!&\!=\!&\!
 \alpha(C(\mathbf{x},t)) \delta_{\sigma_v}(\mathbf{v}\!-\!\mathbf{v}_0) p(\mathbf{x},\mathbf{v},t)  - \Gamma p(\mathbf{x},\mathbf{v},t)\! \int_0^t \!ds \!\int_{\mathbb{R}^2}\! d{\bf v}' p(\mathbf{x},\mathbf{v}'\!,s)  - \mathbf{v}\!\cdot\! \nabla_\mathbf{x}   p(\mathbf{x},\mathbf{v},t)\quad  \nonumber\\
 &\!+\!&\! \beta {\rm div}_\mathbf{v} (\mathbf{v} p(\mathbf{x},\mathbf{v},t))
- {\rm div}_\mathbf{v} \left[\beta\mathbf{F}\left(C(\mathbf{x},t)\right)p(\mathbf{x},\mathbf{v},t)  \right]\! + \frac{\beta}{2}
\Delta_\mathbf{v} p(\mathbf{x},\mathbf{v},t), \label{eq3a}  \\
\frac{\partial}{\partial t}C(\mathbf{x},t) &=& \kappa \Delta_{\mathbf x} C(\mathbf{x},t) - \chi C(\mathbf{x},t) j(\mathbf{x},t), \label{eq3b} \\
p(\mathbf{x},\mathbf{v},0) &=& p_0(\mathbf{x},\mathbf{v}), \quad
C(\mathbf{x},0) =C_0(\mathbf{x}), \label{eq3c}
\end{eqnarray}
where
\begin{eqnarray} 
&&{\alpha(C(\mathbf{x},t))=\frac{A\, C(\mathbf{x},t)}{1+C(\mathbf{x},t)}, \quad
 {\bf F}(C(\mathbf{x},t))= \frac{\delta_1\nabla_{\mathbf x} C(\mathbf{x},t)}{(1+\Gamma_1C(\mathbf{x},t))^{q_1}},\quad \delta_{\sigma_v}(\mathbf{v}-\mathbf{v}_0)= \frac{1}{\pi\sigma_v^2}\, e^{-\frac{|\mathbf{v}-\mathbf{v}_0|^2}{\sigma_v^2}},}
\label{eq3d}\quad \\
&&j(\mathbf{x},t)= \int_{\mathbb R^2} 
\frac{|\mathbf{v}| }{ 1 + e^{(|\mathbf{v}- \mathbf{v}_0 |^2-\eta)/ \epsilon}}
p(\mathbf{x},\mathbf{v},t)\, d \mathbf{v},
\quad \tilde{p}(\mathbf{x},t)= \int_{\mathbb R^2} p(\mathbf{x},\mathbf{v},t)
\, d \mathbf{v}, \label{eq3e}
\end{eqnarray}\end{subequations}  \end{widetext}
for ${\mathbf x} \in \Omega \subset  \mathbb{R}^2$, ${\mathbf v} \in  \mathbb{R}^2$, $t \in [0, \infty).$ The dimensionless parameters $\beta$, $\Gamma$, $\kappa$, $\chi$, $A$, $\Gamma_1$, $\delta_1$, $\eta$, $\epsilon$, $q_1$ and $\sigma_v$ are positive. The integral sink term  $- \Gamma p \int_0^t \tilde{p}(\mathbf{x},s)ds$  (in which $\tilde{p}$ defined in Eq.~\eqref{eq3e} is the marginal tip density) captures the phenomenon that a vessel tip ceases to be active when it encounters another vessel and anastomoses. The anastomosis coefficient $\Gamma$ is calculated by comparison to numerical simulations of the stochastic process, in such a way that the ensemble average of the total number of active tips equals $\int\tilde{p}(\mathbf{x},t) d\mathbf{x}$. The Gaussian function in Eq.~\eqref{eq3d} selects the direction of motion and velocity of new active tips generated by branching. In Eq.~\eqref{eq3b}, TAF diffuses and it is consumed by the flux of advancing tip cells, $j(\mathbf{x},t)$. For the slab geometry of Fig.~\ref{fig:0}, appropriate nonlocal boundary conditions for $p(\mathbf{x},\mathbf{v},t)$ and boundary conditions for $C(\mathbf{x},t)$ are indicated in \cite{bon18}. There an appropriate explicit finite-difference numerical scheme is described, and its stability and convergence are proved. Numerical simulations of the PDEs illustrating the formation of a soliton solution and comparison with the solution of the stochastic model can also be found in \cite{bon18}. Global existence, uniqueness and well-posedness results for Equations \eqref{eq3} can be found in \cite{car16,car17}. 

In the overdamped limit (small inertia), it is possible to obtain a simpler equation for the density of active tip cells $\tilde{p}(\mathbf{x},t)$ \cite{bon16,bon16pre} 
\begin{subequations}\label{eq4}
\begin{eqnarray}
\frac{\partial\tilde{p}}{\partial t} & + & \nabla\cdot(\mathbf{F}\,\tilde{p})-\frac{1}{2\beta}\nabla^2\tilde{p} \,=\, \tilde{\mu}\,\tilde{p}\nonumber\\
&& -\Gamma\,\tilde{p}\int_0^t\tilde{p}(\mathbf{x},s)\,ds, \label{eq4a}\\
\frac{\partial C}{\partial t} & = & \kappa \nabla^2 C - \tau C\,\tilde{p}, \label{eq4b}\\
\mathbf{F} & = & (F_x,F_y) = \frac{\delta_1}{\beta}\,\frac{\nabla C}{1+\Gamma_1 C}, \label{eq4c}
\end{eqnarray}
\end{subequations}
where $\tau=|\mathbf{v}_0|\chi$, we have set $q_1=1$, and $\tilde{\mu}$ is a function of $C$ related to branching (see \cite{bon16pre}). These equations need to be supplemented with initial and boundary conditions appropriate for the configuration that we study. Similarly to previous works \cite{cap09,bon14,ter16,bon16,bon16pre}, we consider a strip geometry with a vertical primary vessel at $x=0$ and a TAF source located at $x=L_x$, as sketched in Fig.~\ref{fig:0}. Note that $\tilde{p}(\mathbf{x},t)$ and $\int_0^t\tilde{p}(\mathbf{x},s) ds$ in Eq.~\eqref{eq4a} correspond to $N(\mathbf{x},t)$ and $E(\mathbf{x},t)$ in Eqs.~\eqref{eq1}, provided tip-to-tip anastomosis is ignored ($a_n=0$) and there are no stalk cells initially ($H(x)=0$ in Eq.~\eqref{eq2d}). 

After a transient stage, the density of active tips $\tilde{p}(\mathbf{x},t)$
evolves to a soliton-like wave with slowly varying velocity and size, which we can describe by a combination of asymptotics and numerical simulations~\cite{bon16,bon16pre}. This stage ends when the soliton approaches the tumor at $x=L_x$.


\section{Soliton description} \label{sec:3}
In order to characterize the evolution of the tip cell density $N$, we seek a wavelike approximation
for the LO-PDEs in Eqs.~\eqref{eq1} and \eqref{eq2}. Following the discussion in
\cite{bon16pre}, the soliton has the form
\beqn
N_s(x,t) \!=\! \frac{(2Ka_e+\tilde{\mu}^2)c}{2a_e(c-F_x)}\mbox{sech}^2\!\left(\! \frac{\sqrt{2Ka_e+\tilde{\mu}^2}}{2(c-F_x)}
\xi\! \right)\!, \label{eq5}
\eeqn
where
\beqn
F_x  = \chi \,\frac{\partial C}{\partial x}\,, \quad 
\tilde{\mu} = \lambda \,\frac{C}{1 + C}\,, \quad  \xi = x - X\,. \label{eq6}
\eeqn
Note that, when $C$ varies slowly in time and space, $F_x$ and $\tilde{\mu}$ in Eq.~\eqref{eq6} are suitable average values (see below). On the other hand, $K(t)$, $c(t)$ and $X(t)$ are time-dependent \textit{collective coordinates} describing the shape and velocity of the soliton. They will be computed by integrating a system of three coupled ordinary differential equations.
Thus, Eq.~\eqref{eq5} yields a slowly varying soliton-like approximation
of the tip cell density, which is valid after a formation stage and far away from the
boundary $x = L_x$ where the TAF source is located (see \cite{bon16pre}). \\

Now, following the illustration given in \cite{bon16pre}, we shall deduce the system of collective coordinate equations (CCEs) for the LO-PDEs under the assumptions of small diffusion and a TAF concentration that varies slowly in space and time. First, we observe that $N_s$ is a function of $\xi$, and the space and time variables through $C$, namely
\begin{equation}
	N_s \,=\, N_s\!\left(\xi;K,c,\tilde{\mu}(C),F_x\!\left(\frac{\partial C}{\partial x}\right)\!\right)\!. 
	\label{eq7}
\end{equation}
We assume that the TAF variations over time and space produce terms that are small compared to $\partial N_s/\partial \xi$.
In addition, we suppose that $\tilde{\mu}(C)$ is approximately constant (since $C$ is slowly varying)
and ignore $\partial^2 N_s/\partial i \partial j$ for $i,j = K,\,F_x$.
Then, plugging Eq.~\eqref{eq5} into Eq.~\eqref{eq1a}, noting that Eq.~\eqref{eq1a} has a soliton solution
for zero-diffusion and constant $\nabla C$, and taking Eq.~\eqref{eq7} into account, we obtain
\begin{equation}
\frac{\partial N_s}{\partial K} \dot{K} + \frac{\partial N_s}{\partial c} \dot{c} \,=\, \mathcal{A}, 
\label{eq8}
\end{equation}
where
\begin{widetext}
\begin{equation}
\mathcal{A} \,=\, \! D\frac{\partial^2 N_s}{\partial\xi^2}\! - \! N_s\nabla\!\cdot\!\mathbf{F}\!-\!\frac{\partial N_s}{\partial F_x}\!\bigg[ \mathbf{F}\!\cdot\!\nabla F_x\!-\!D\nabla^2 F_x \bigg]\!
+ 2D\frac{\partial^2 N_s}{\partial \xi \partial F_x}\frac{\partial F_x}{\partial x} - \mu\,a_n N_s^2, \label{eq9}
\end{equation}
\end{widetext}
with $\mathbf{F} = \chi \nabla C$. Indeed, all terms in Eq.~\eqref{eq9}
depending on $\mathbf{F}$ or $F_x$ (hence on $C$) are reduced to the spatial variable $x$ by averaging in the $y$-direction,
as mentioned before (see \cite{mar21}).
Next, we multiply Eq.~\eqref{eq8} by $\partial N_s/\partial K$ and integrate over $x$.
We consider a fully formed soliton, far from the primary vessel and the TAF source.
As it exponentially decays for $|\xi |\gg 1$,
the soliton is regarded to be localized on some finite interval $(-\mathcal{L}/2,\mathcal{L}/2)$,
where the TAF varies slowly. Therefore, we can approximate \cite{bon16pre}
\begin{eqnarray}
&&	\!\int_\mathcal{I}\! \phi(N_s(\xi;x,t),x) dx\nonumber\\
&&	\approx\frac{1}{\mathcal{L}}\int_\mathcal{I}\,\!\!\left(\int_{-\mathcal{L}/2}^{\mathcal{L}/2}\! \phi(N_s(\xi;x,t),x) d\xi\!\right)\! dx, \label{eq10}
\end{eqnarray}
where the interval $\mathcal{I}$ has extension equal to $\mathcal{L}$ and should contain most of the soliton
(here $\phi$ is a generic function of $N_s$ and $x$).
Thus, the CCEs only hold after an initial soliton formation stage and far from the TAF source, regions that must be
excluded from $\mathcal{I}$.
Similarly, we multiply Eq.~\eqref{eq8} by $\partial N_{s}/\partial c$ and integrate over $x$.
From the two resulting formulas, we then find $\dot{K}$ and $\dot{c}$. Since the factor $1/\mathcal{L}$ cancels out
and the soliton tails decay to zero, we can set $\mathcal{L}\to\infty$ and obtain the following CCEs \cite{bon16pre}:
\begin{widetext}
\begin{subequations}\label{eq11}
\begin{eqnarray}
&&	\dot{K} = \frac{\int_{-\infty}^\infty \frac{\partial N_s}{\partial K}\mathcal{A}\,d\xi \int_{-\infty}^\infty\!\!\left(\frac{\partial N_s}{\partial c}\right)^2\!\!d\xi\! \,- \int_{-\infty}^\infty \frac{\partial N_s}{\partial c}\mathcal{A}\,d\xi \int_{-\infty}^\infty\!\frac{\partial N_s}{\partial K}\frac{\partial N_s}{\partial c} d\xi\!}{\int_{-\infty}^\infty\!\!\left(\frac{\partial N_s}{\partial K}\right)^2\!\!d\xi\!\,\int_{-\infty}^\infty\!\!\left(\frac{\partial N_s}{\partial c}\right)^2\!\!d\xi\!\,-\left(\int_{-\infty}^\infty \frac{\partial N_s}{\partial c}\frac{\partial N_s}{\partial K}d\xi\right)^2}, \label{eq11a}\\
&&	\dot{c} = \frac{\int_{-\infty}^\infty \frac{\partial N_s}{\partial c}\mathcal{A}\,d\xi\int_{-\infty}^\infty\!\!\left(\frac{\partial N_s}{\partial K}\right)^2\!\!d\xi\!\,- \int_{-\infty}^\infty \frac{\partial N_s}{\partial K}\mathcal{A}\,d\xi\int_{-\infty}^\infty\!\frac{\partial N_s}{\partial K}\frac{\partial N_s}{\partial c} d\xi\!}{\int_{-\infty}^\infty\!\!\left(\frac{\partial N_s}{\partial K}\right)^2\!\!d\xi\!\,\int_{-\infty}^\infty\!\!\left(\frac{\partial N_s}{\partial c}\right)^2\!\!d\xi\!\,-\left(\int_{-\infty}^\infty \frac{\partial N_s}{\partial c}\frac{\partial N_s}{\partial K}d\xi\right)^2}, \label{eq11b}
\end{eqnarray}
together with
\begin{equation}
	\dot{X} = c\,. \label{eq11c}
\end{equation}
\end{subequations}\end{widetext}

In these equations, all terms depending on $C$ that vary slowly with $x$ are averaged over the interval $\mathcal{I}$, which will be specified in sections devoted to the numerical results. On the other hand, the penultimate term in Eq.~\eqref{eq9} is odd in $\xi$ and does not contribute to the integrals in Eqs.~\eqref{eq11a}--\eqref{eq11b}. Most of these integrals have been calculated in Appendix D in \cite{bon16pre}. The only two new integrals correspond to the last term in Eq.~\eqref{eq9}, which models tip-to-tip anastomosis, and they are
\begin{equation}
\int_{-\infty}^\infty\!\!\,\frac{\partial N_s}{\partial K}\,N_s^2 \!\,\,d\xi \!\,\,=\,\,\! \frac{4 c^3(2K a_e+\tilde{\mu}^2)^\frac{3}{2}}{9a_e^2(c-F_x)^2}, \label{eq12}
\end{equation}
\begin{eqnarray}
\int_{-\infty}^\infty\!\!\frac{\partial N_s}{\partial c}N_s^2\!d\xi \!=\! \frac{4c^2\!(c\!-\!3F_x)(2K\! a_e\!+\!\tilde{\mu}^2)^\frac{5}{2}}{45a_e^3(c-F_x)^3}\!.\quad \label{eq13}
\end{eqnarray}
These integrals will be relevant to the analysis reported in Section \ref{sec:7}.


\section{One-dimensional linear TAF} \label{sec:4}
We first assume, as in Figure \ref{fig:1}, that a quasi-steady, 1D linear TAF concentration
\beqn
 C(x) = x, \label{eq14}
\eeqn
for $0 < x < 20$, drives the dynamics. Then, we seek a soliton approximation for the tip cell density $N$ as illustrated in Section \ref{sec:3}. The model parameters are set to the values indicated in Section \ref{sec:2}. Moreover, the initial conditions for the CCEs in Eqs.~\eqref{eq11} are given at $t_0 = 10$ (estimated as the soliton formation stage) as follows: $X(t_0)$ is the location of the maximum of $N$ at $t_0$, $c(t_0) = X(t_0)/t_0$, and $K(t_0)$ is determined so that the soliton peak coincides with the maximum tip cell density at $t_0$. After solving the CCEs, the soliton in Eq.~\eqref{eq5} is reconstructed and compared to the dynamics of $N$ given by the numerical solutions of Eqs.~\eqref{eq2}. \\

\begin{figure}[h!]
	\begin{center}
		\includegraphics[width=8cm,height=4cm]{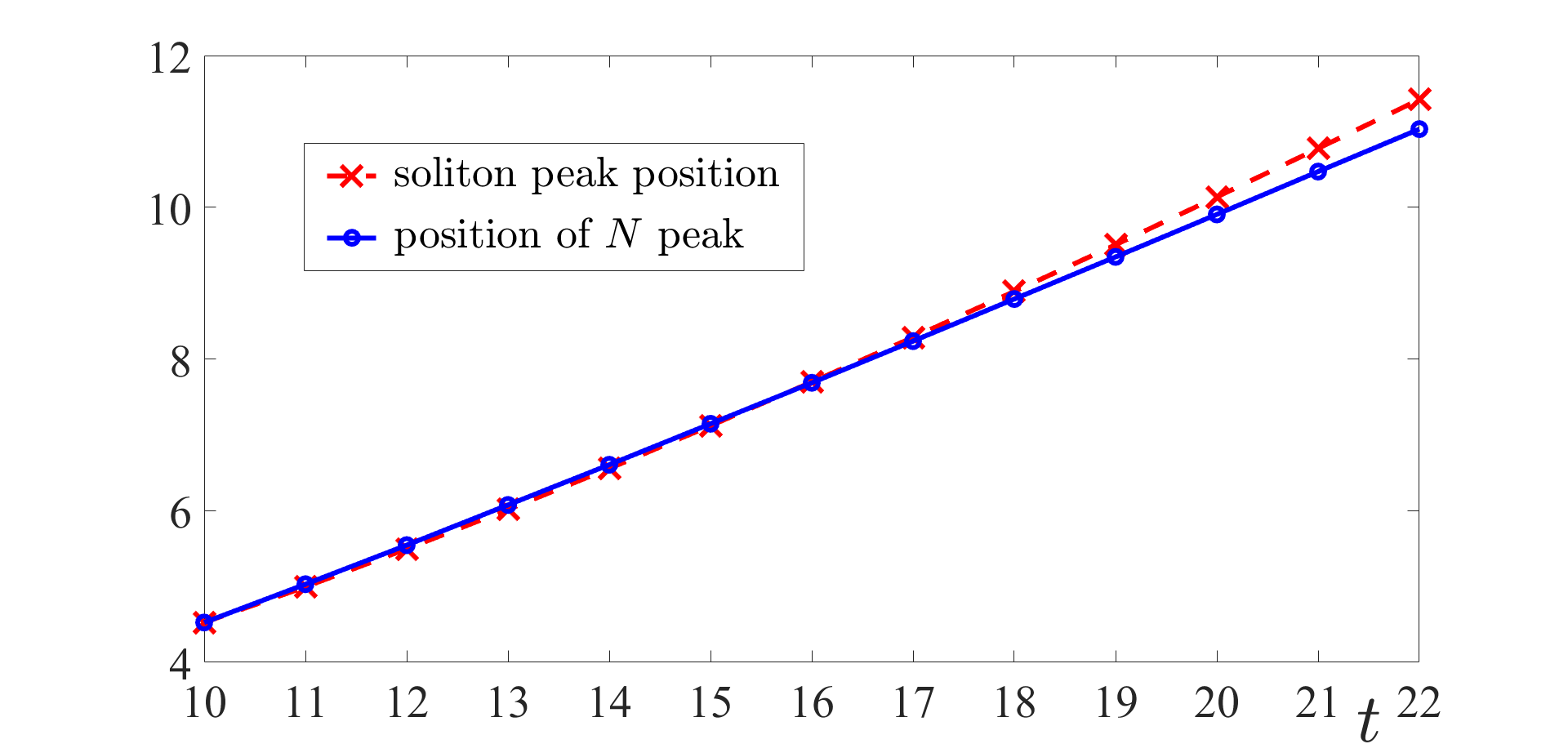} \\[2mm]
		\includegraphics[width=8cm,height=4cm]{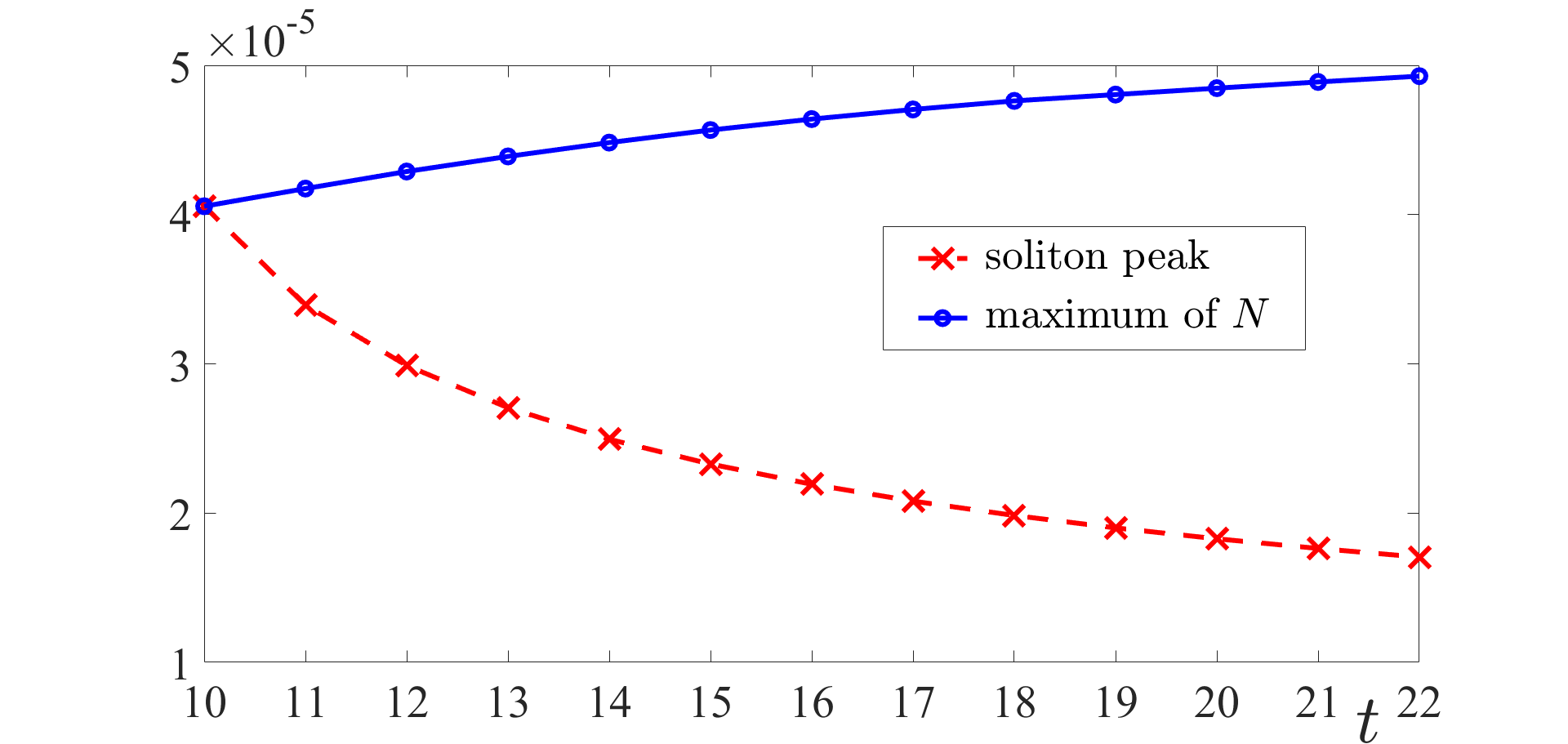}
	\end{center}
	\vskip-0.5cm
	\caption{Time evolution of position (top) and value (bottom) of the maximum tip cell density $N$ as computed
	by solving Eqs.~\eqref{eq2} (solid, blue line) and the soliton in Eq.~\eqref{eq5} (dashed, red line),
	for the TAF concentration in Eq.~\eqref{eq14} and $L_x = 20$. \label{fig:2}}
\end{figure}

Figure \ref{fig:2} shows the position and value of the peak (i.e., the maximum) tip cell density as computed by numerical simulations of Eqs.~\eqref{eq2} and the soliton in Eq.~\eqref{eq5}. We can observe that, while the peak location is well predicted in the time interval $10 \leq t \leq 22$, the approximation of the maximum value of $N$ fails. This is due to the simple form of the TAF concentration in Eq.~\eqref{eq14}. If $C=x$, then the terms in Eq.~\eqref{eq9} that depend on advection (i.e., differentials of $\mathbf{F}$) vanish and do not contribute to the CCEs. However, these terms are crucial for the correct evaluation of the maximum tip density $N$, which explains the discrepancy shown in Fig.~\ref{fig:2}.
%
%


\section{Two-dimensional Gaussian TAF} \label{sec:5}
In order to improve the soliton approximation, we now assume that the quasi-steady TAF concentration
is a 2D Gaussian \cite{bon16pre}:
\beqn
 C(x,y) = a\, e^{-(x-21)^2/\sigma_x^2 -(y-0.5)^2/\sigma_y^2}, \label{eq15}
\eeqn
 for $0 < x < 20$ and $0 < y < 1$, with $a = 30$, $\sigma_x = 15$, and $\sigma_y = 4$. 
These values are chosen in such a way that the generated soliton wave has a similar velocity and height as the wave generated for the linear profile of Eq.~\eqref{eq14}. \\

\begin{figure}[h!]
	\begin{center}
		\includegraphics[width=8cm]{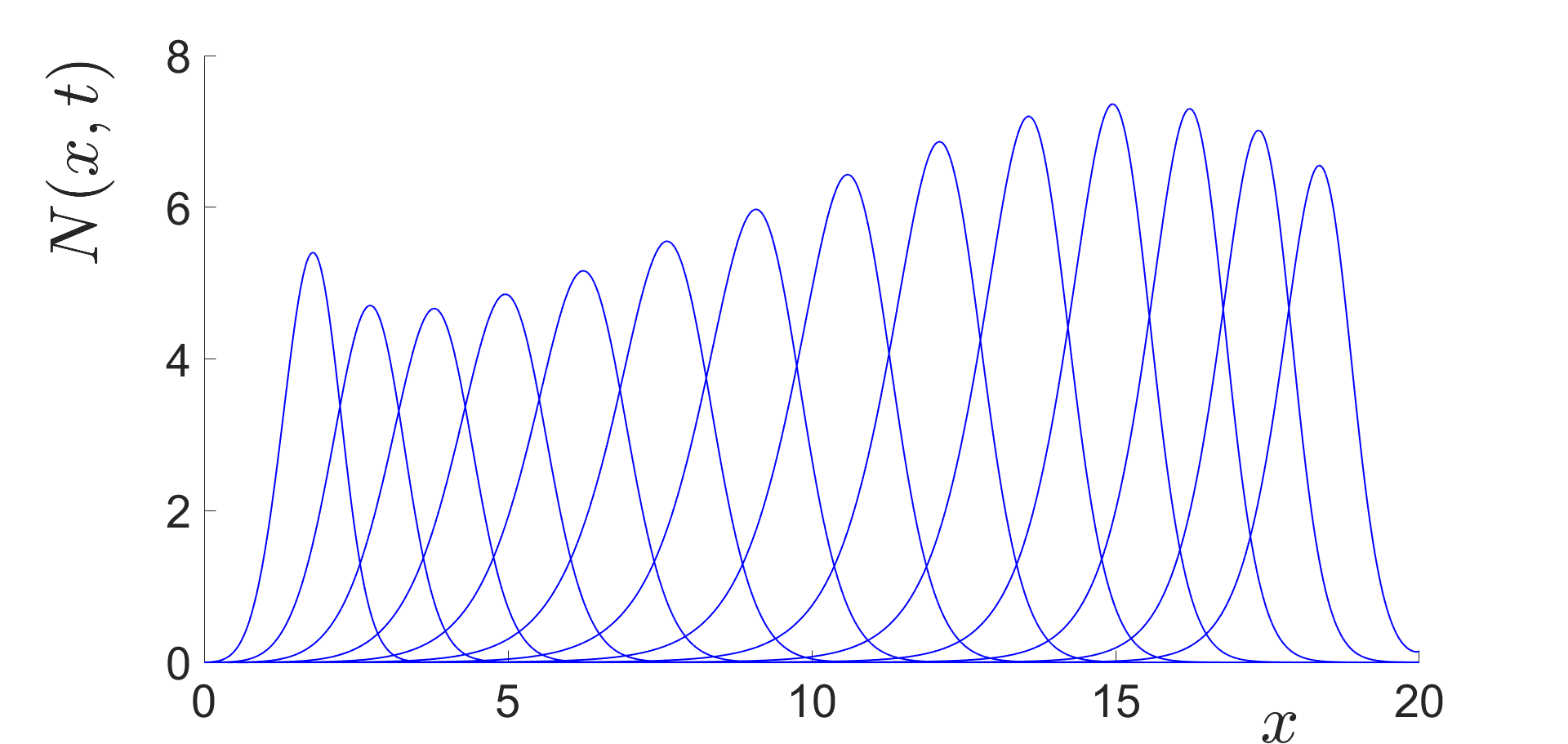}
	\end{center}
	\vskip-0.5cm
	\caption{Counterpart of Figure \ref{fig:1}, for the (column-averaged) TAF concentration in Eq.~\eqref{eq15}.
	\label{fig:3}}
\end{figure}

\begin{figure}[h!]
	\begin{center}
		\includegraphics[width=8cm,height=4cm]{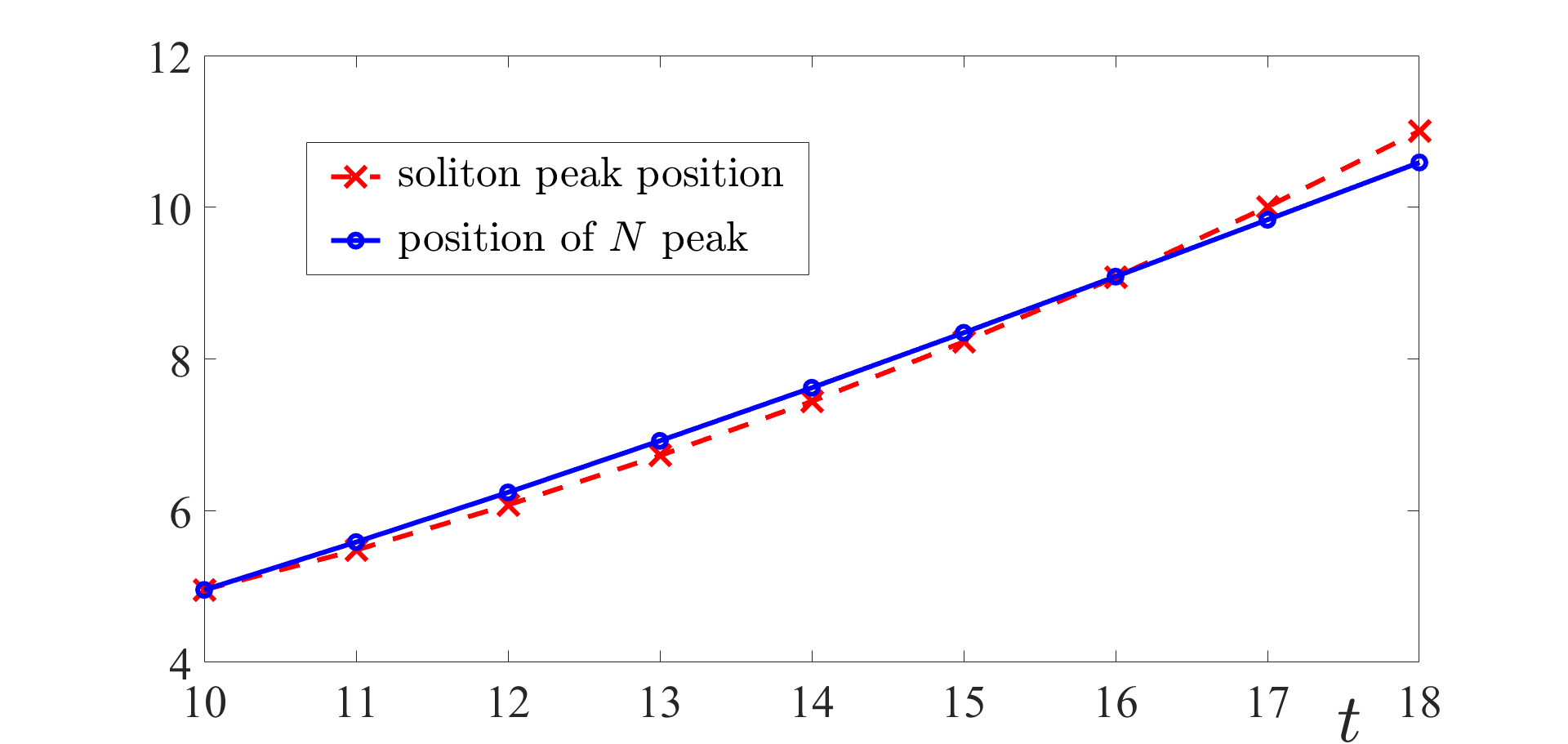} \\[2mm]
		\includegraphics[width=8cm,height=4cm]{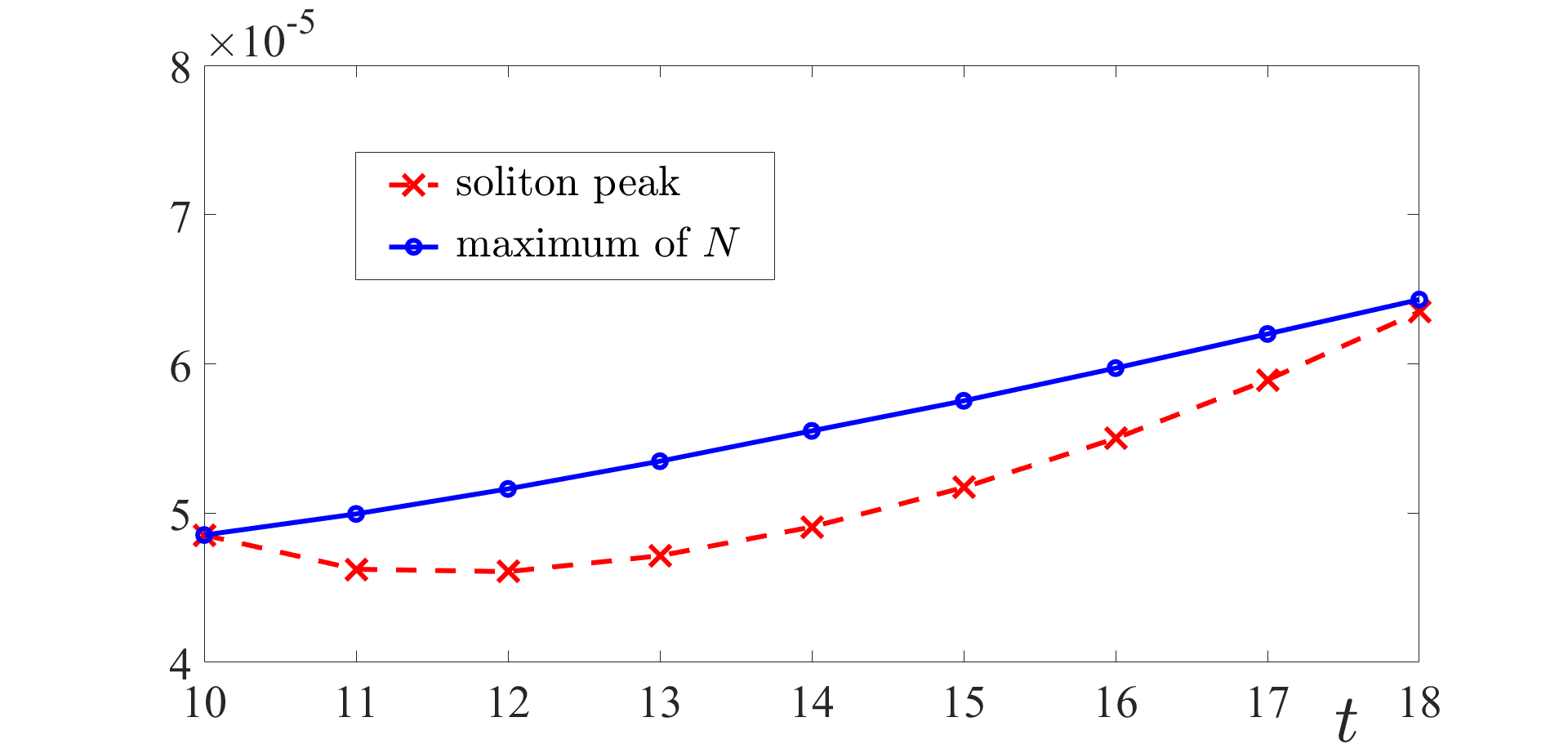}
	\end{center}
	\vskip-0.5cm
	\caption{Counterpart of Figure \ref{fig:2}, for the (column-averaged) TAF concentration in Eq.~\eqref{eq15}. \label{fig:4}}
\end{figure}

In Eqs.~\eqref{eq9} and \eqref{eq11}-\eqref{eq13} there are terms depending on the 2D TAF concentration. Firstly, we calculate them as functions of $x$ and $y$ on the 2D domain. Secondly, we column-average them to eliminate their dependence on $y$. Next, we average over $x\in \mathcal{I} = (0,2]$ all terms in the CCEs that depend on the TAF and we set the initial conditions for the CCEs at $t_0 = 10$. As before, the model parameter values are those indicated in Section \ref{sec:2}. The counterparts of Figures \ref{fig:1} and \ref{fig:2} are given by Figures \ref{fig:3} and \ref{fig:4}, respectively. Now, both location and value of the peak tip cell density are well predicted by the soliton in Eq.~\eqref{eq5} over the time interval $10 \leq t \leq 18$. Indeed, the TAF concentration in Eq.~\eqref{eq15} contributes to all relevant terms
in the system of CCEs; cf. Eq.~\eqref{eq9}. It is worth remarking that the soliton approximation is robust with respect to changing the values of $a$, $\sigma_x$, and $\sigma_y$ in Eq.~\eqref{eq15}. Finally, we observe that the soliton description is limited to a finite time window: its validity is affected by the no-flux boundary condition for the tip cell density imposed in the model at $x = L_x = 20$.
%
%


\section{Two-dimensional TAF on a larger spatial domain} \label{sec:6}
\begin{figure}[h!]
	\begin{center}
		\includegraphics[width=8cm]{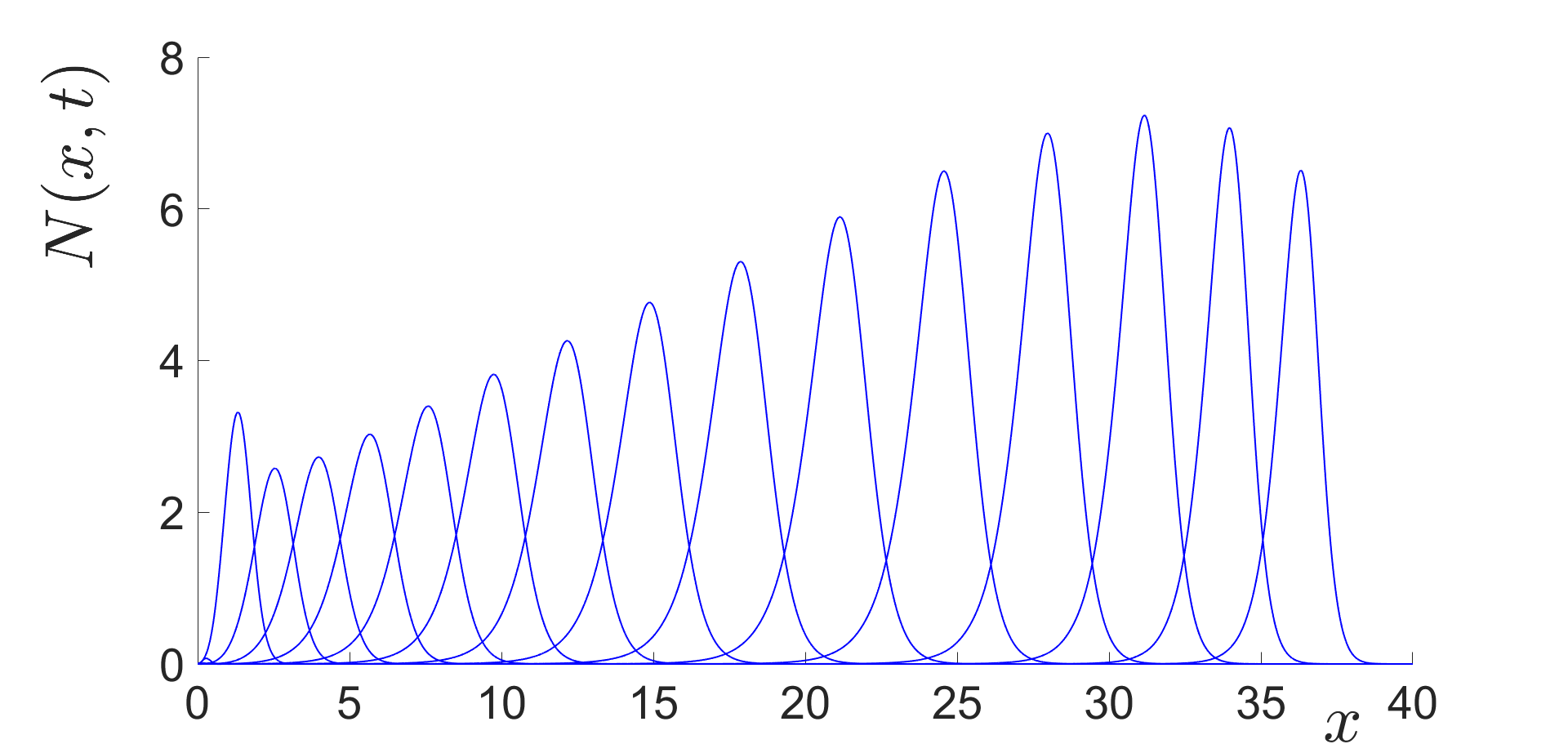}
	\end{center}
	\vskip-0.5cm
	\caption{Snapshots (at distance $\Delta t = 4$) of the time evolution between $t = 4$ and $t = 60$
		of the tip cell density as numerically
		computed from Eqs.~\eqref{eq2}, for the (column-averaged)
		TAF concentration in Eq.~\eqref{eq16} and $L_x = 40$. The scale on the vertical axis
		is $\times 10^{-5}$. \label{fig:5}}
\end{figure}

In this section, we study how the no-flux boundary condition imposed on the tip cell density at $x = L_x$ affects the soliton approximation. Hence, we consider a quasi-steady, 2D Gaussian TAF concentration on a larger spatial domain, namely
\beqn
 C(x,y) = a\, e^{-(x-41)^2/\sigma_x^2 -(y-0.5)^2/\sigma_y^2}, \label{eq16}
\eeqn
for $0 < x < 40$ and $0 < y < 1$, with $a = 50$, $\sigma_x = 23$, and $\sigma_y = 4$.
For the sake of comparison with Section~\ref{sec:5}, the parameter values in Eq.~\eqref{eq16} have been selected so as to generate (via numerical simulations of Eqs.~\eqref{eq2}) a tip cell density that reaches its tallest maximum at around $75\%$ of the spatial domain
in approximately twice the time as in Section \ref{sec:5}; compare Figures \ref{fig:3} and \ref{fig:5}. \\

\begin{figure}[h!]
	\begin{center}
		\includegraphics[width=8cm,height=4cm]{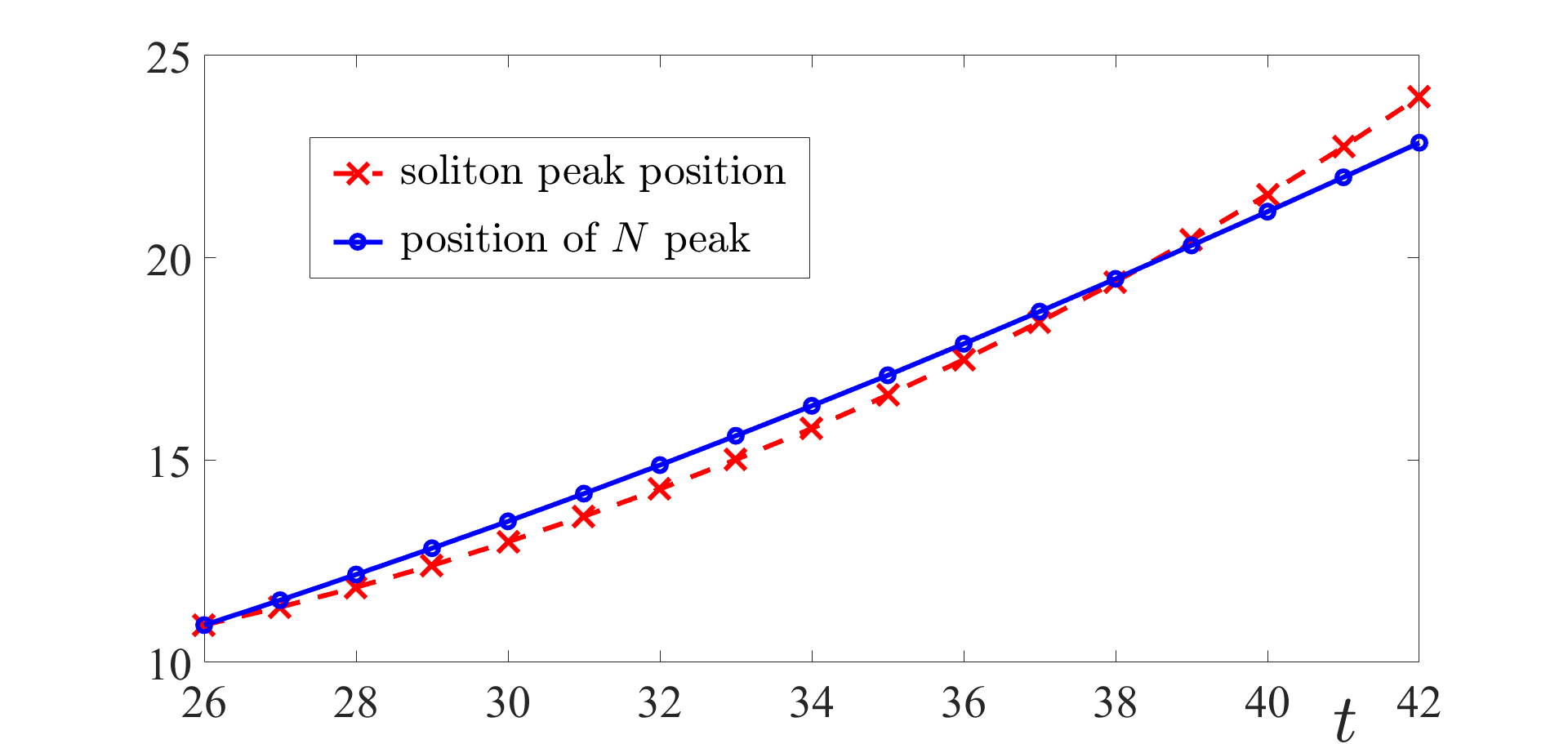} \\[2mm]
		\includegraphics[width=8cm,height=4cm]{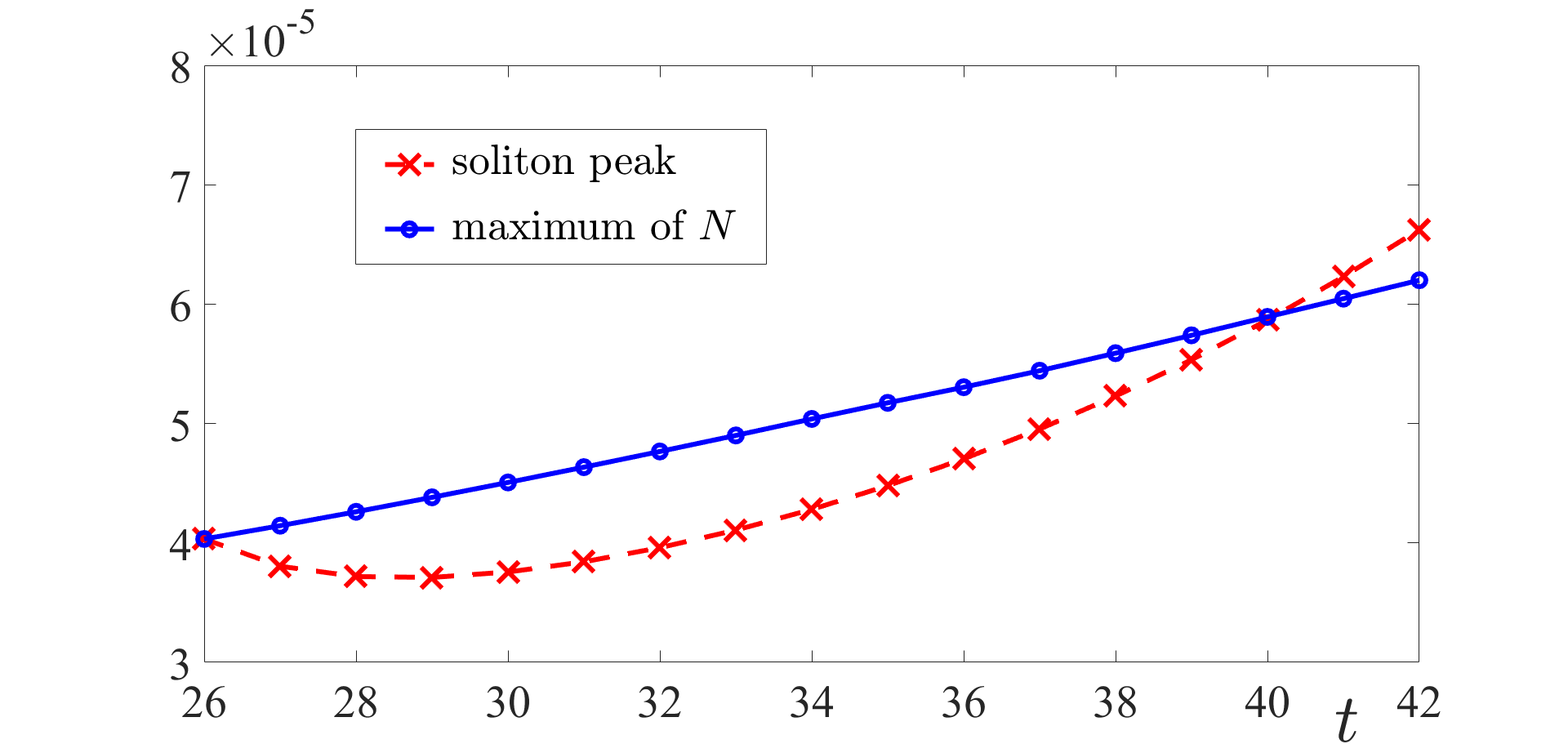}
	\end{center}
	\vskip-0.5cm
	\caption{Counterpart of Figure \ref{fig:2}, for the (column-averaged) TAF concentration in Eq.~\eqref{eq16}
	and $L_x = 40$. \label{fig:6}}
\end{figure}

Now, the $x$-averages shall be computed in the spatial interval $\mathcal{I} = (0,4]$,
even though restricting the averaging to a small subinterval (e.g., $\mathcal{I} = [3.2,3.3]$)
may result in a slightly better outcome.
The model parameters are again set to the values indicated in Section \ref{sec:2}.
After an initial stage of $t_0 = 26$ (when the tip cell density reaches its peak at around
$25\%$ of the spatial domain, as in the case of Section \ref{sec:5}), the soliton in Eq.~\eqref{eq5}
is able to correctly predict both location and value of the maximum of $N$
on the time interval $26 \leq t \leq 42$. Indeed, the soliton takes longer
to reach a point where the effect of the no-flux boundary condition (imposed at $x = 40$
on the tip cell density) starts playing a role.
Thus, the soliton approximation holds over a much wider time window, as illustrated in Figure \ref{fig:6}.
%
%


\section{Effect of tip-to-tip anastomosis} \label{sec:7}
\begin{figure}[h!]
	\begin{center}
		\includegraphics[width=8cm]{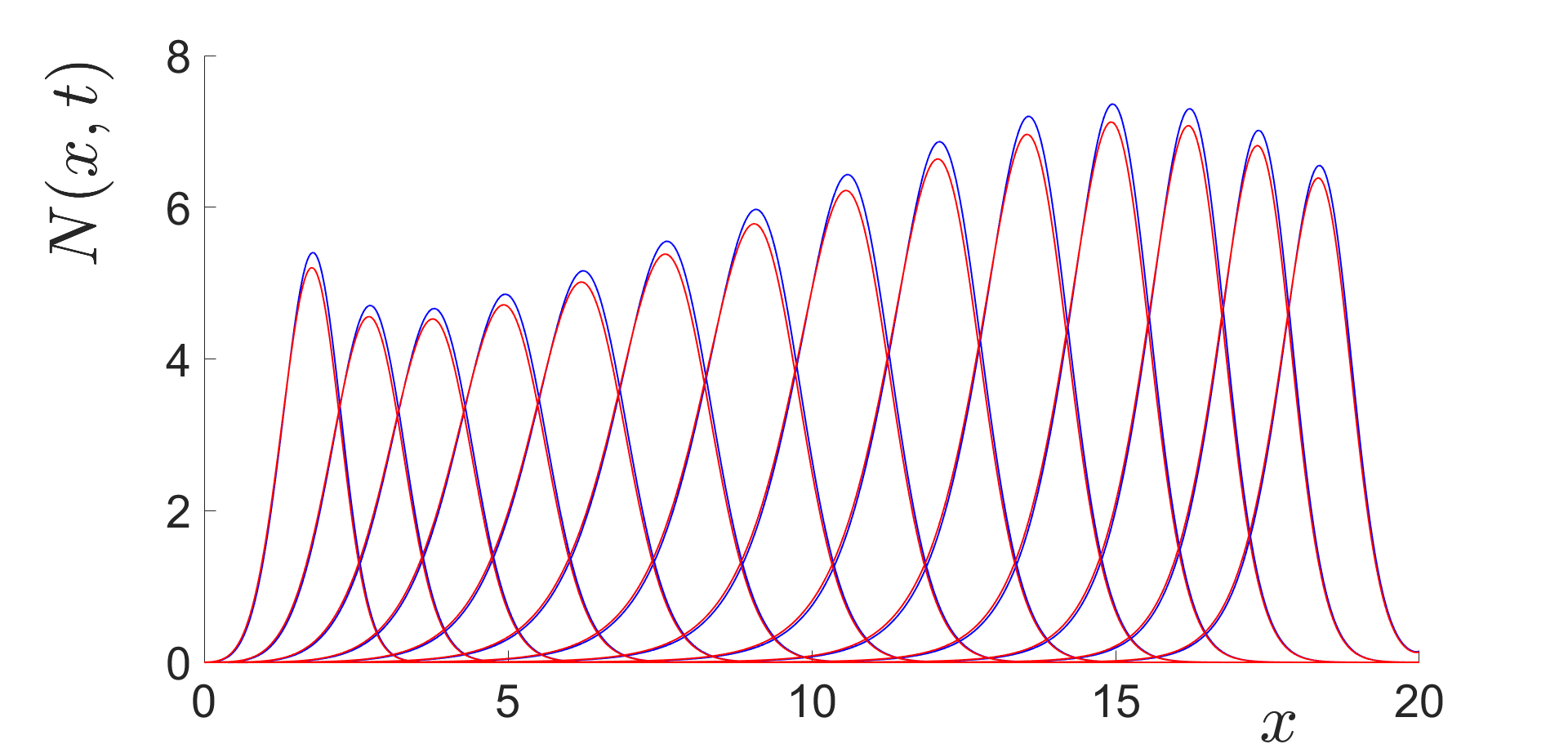}
	\end{center}
	\vskip-0.5cm
	\caption{Counterpart of Figure \ref{fig:1}, for the TAF concentration in Eq.~\eqref{eq15}, considering
		$a_n = 0$ (blue lines) and $a_n = 1$ (red lines). \label{fig:7}}
\end{figure}

Let us now model tip-to-tip anastomosis by considering $a_n \neq 0$ in Eqs.~\eqref{eq2}.
Figure \ref{fig:7} shows the tip cell density evolution over time for the same parameter values as indicated
in Section \ref{sec:2}, with ($a_n = 1$) and without ($a_n = 0$) tip-to-tip anastomosis,
for the TAF concentration in Eq.~\eqref{eq15}.
Indeed, only a small difference can be appreciated in the overall advance of the wavelike profiles between the two cases.
Calibrating the value of the parameter $a_n > 0$ allows modulation of the intensity of this mechanism
of vessel fusion. \\

\begin{figure}[h!]
	\begin{center}
		\includegraphics[width=8cm,height=4cm]{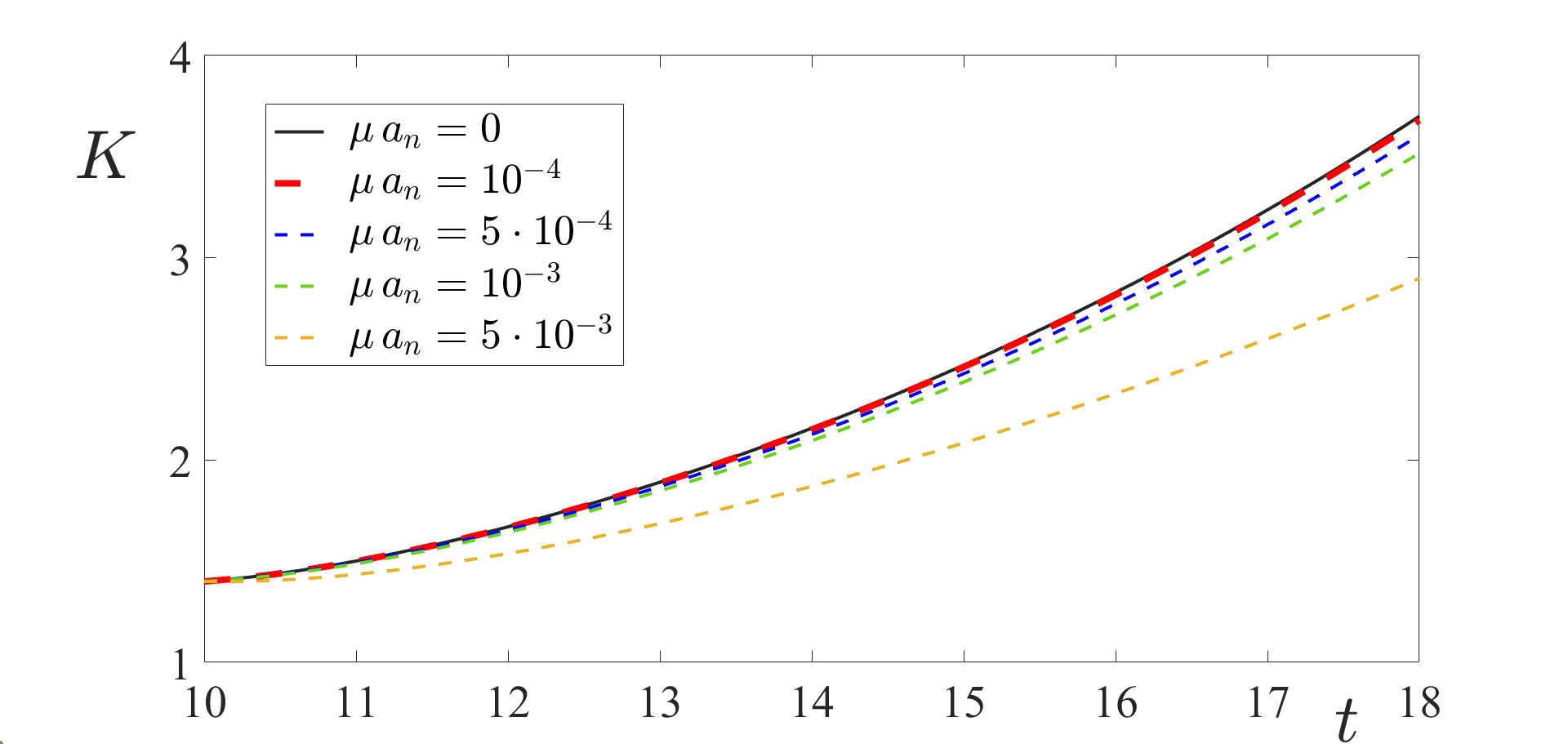}
		\includegraphics[width=8cm,height=4cm]{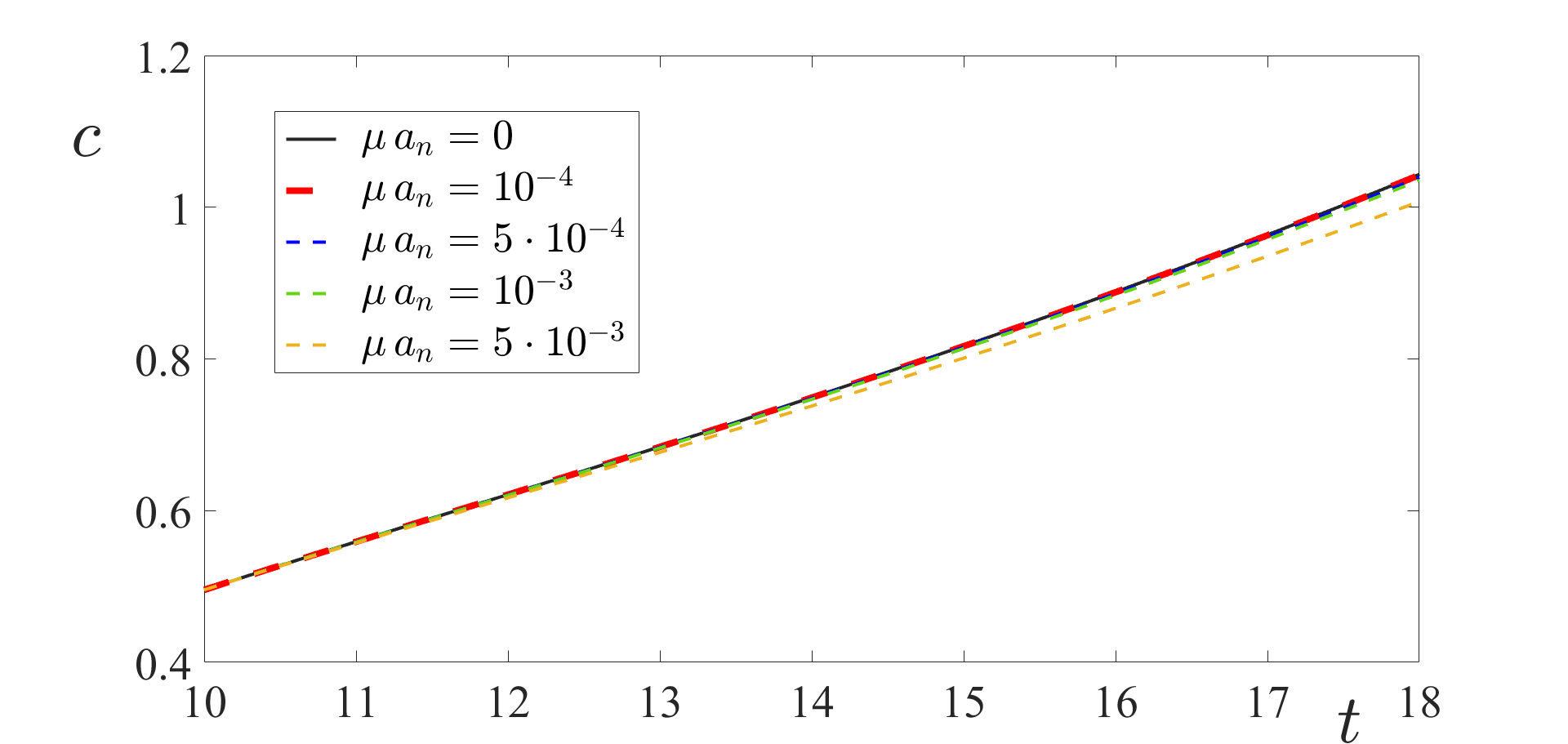}
		\includegraphics[width=8cm,height=4cm]{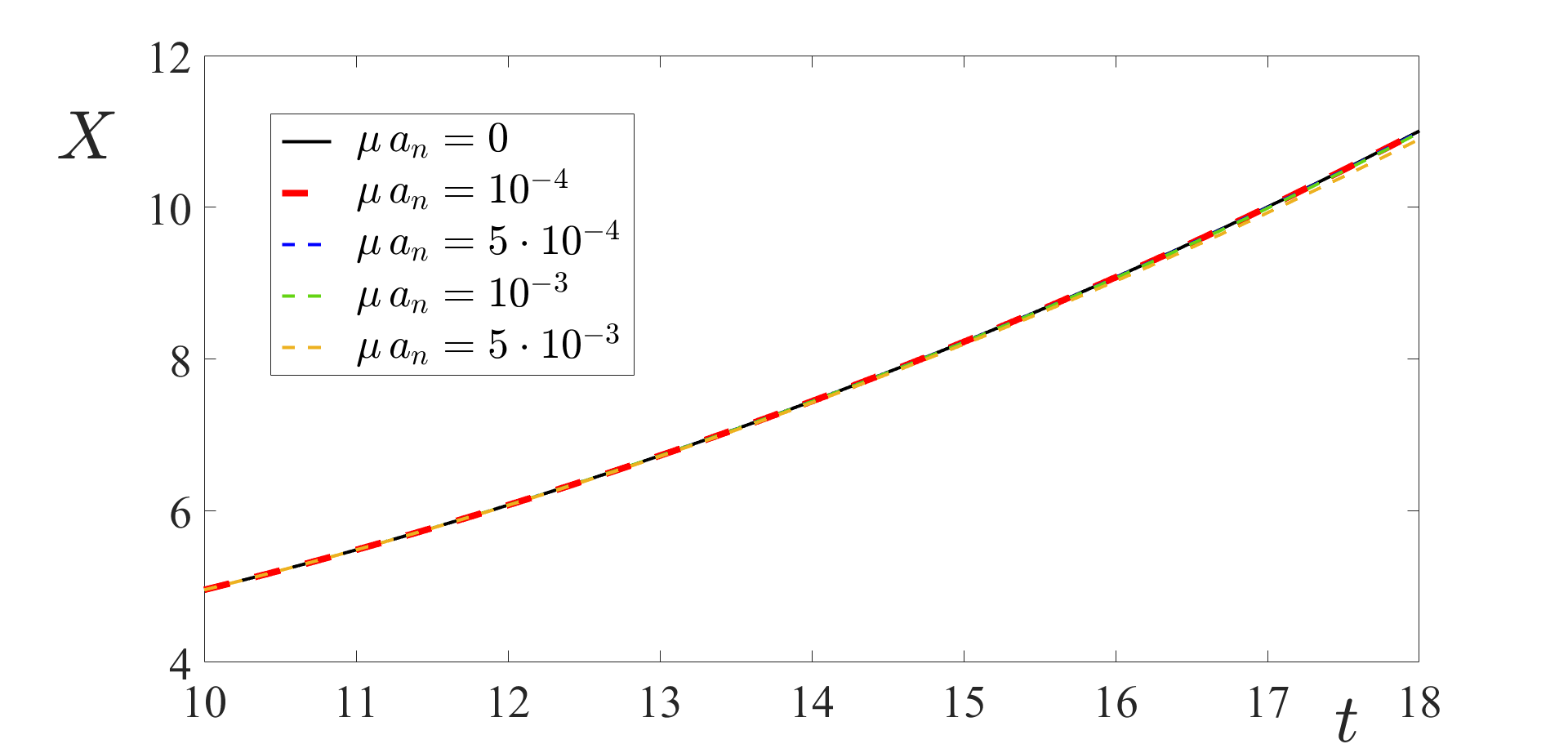}
	\end{center}
	\vskip-0.5cm
	\caption{Temporal evolution of the three collective coordinates, $K(t)$, $c(t)$, and $X(t)$,
	for the TAF concentration in Eq.~\eqref{eq15} and different values of $\mu\,a_n$. \label{fig:8}}
\end{figure}

The effect on the soliton approximation can be quantified by taking into account the term
$-\mu \,a_n N_s^2$ in Eq.~\eqref{eq9}, which modifies the CCEs according to the integrals
in Eqs.~\eqref{eq12}--\eqref{eq13}. Considering $x$-averages of the TAF-dependent terms on the spatial interval $\mathcal{I} = (0,2]$ and initial conditions at $t_0 = 10$, the system of CCEs is integrated in the interval $10 \leq t \leq 18$ for different values
of $\mu\,a_n$ (here, $\mu = 236$). Figure \ref{fig:8} illustrates the resulting temporal behavior of the three collective
coordinates, $K(t)$, $c(t)$ and $X(t)$, in comparison with their evolution for no tip-to-tip anastomosis
($a_n = 0$). We can note that, in the simulated cases, the influence on the overall propagation
velocity of the soliton is fairly small. As a consequence, the location of its peak remains
unaffected. Undoubtedly, the shape coordinate $K$ is the most sensitive to the presence of the
new mechanism. Indeed, $\mu\,a_n < 5 \cdot 10^{-3}$ should be considered
in order to preserve a soliton description of the tip cell density within an accuracy similar to
the case without tip-to-tip anastomosis (see Section \ref{sec:5}).

%
%


\section{Concluding remarks}\label{sec:8}
We have found that the leading order dynamics of ST-PDE and P-PDE systems evolves to a quasi-1D soliton-wave after a transient formation stage. The velocity and shape of the soliton are well approximated by CCEs until it approaches the tumor at $x=L_x$. The system of CCEs gives an accurate representation of the soliton shape and motion for a quasi-steady Gaussian TAF concentration, whereas the shape is not correctly described if the TAF profile is purely linear. However, we should recall that the 1D LO-PDE is the result of averaging the corresponding 2D PDE over the transversal coordinate. When we consider a 2D TAF profile and average it over the transversal coordinate, the soliton wave describes well the velocity and shape of the evolving system of blood vessels.

The LO-PDEs are analogous to the overdamped limit of the continuum equation for the density of active tip cells corresponding to the hybrid stochastic angiogenesis model of \cite{bon14,ter16} (which does not include tip-to-tip anastomosis). Thus, the soliton seems to be an attractor for a class of continuum equations resulting from coarse-graining different discrete and stochastic angiogenesis models. Provided external fields informing chemotaxis, haptotaxis \cite{cap09}, and so on, evolve slowly over longer spatial scales, we can consider their effects by appropriately modifying the CCEs of the soliton \cite{bon17}. To extend the analysis of the LO-PDE, we should model the tumor that emits TAF and study its interaction with the arriving soliton wave. This is outside the scope of the present paper.

\acknowledgements
We acknowledge fruitful discussions with H. M. Byrne. This work has been supported by the FEDER~/~Ministerio de Ciencia, Innovaci\'on y Universidades -- Agencia Estatal de Investigaci\'on grant PID2020--112796RB--C22, by the Madrid Government (Comunidad de Madrid, Spain) under the Multiannual Agreement with UC3M in the line of Excellence of University Professors (EPUC3M23), and in the context of the V PRICIT (Regional Programme of Research and Technological Innovation). WDM acknowledges support from the Keasbey Memorial Foundation, the University of Oxford (postgraduate scholarship), and the Advanced Grant Nonlocal-CPD (Nonlocal PDEs for Complex Particle Dynamics: Phase Transitions, Patterns and Synchronization) of the European Research Council Executive Agency (ERC) under the European Union's Horizon 2020 research and innovation programme (grant agreement No. 883363). PKM and WDM would like to thank the Isaac Newton Institute for Mathematical Sciences, Cambridge, for support and hospitality during the programme Mathematics of Movement where work on this paper was undertaken. This work was supported by EPSRC grant no EP/R014604/1.


\end{document}